# A double-continuum transport model for segregated porous media: derivation and sensitivity analysis-driven calibration

G. Ceriotti[a], A. Russian[a], D. Bolster[b], G. Porta[a]

[a] Dipartimento Ingegneria Civile ed Ambientale Politecnico di Milano, Piazza L. Da Vinci, 32, Milano I-20133 , Italy

[b] Dept. of Civil and Environmental Engineering and Earth Sciences, University of Notre Dame, IN, USA

Abstract

We present and derive a novel double-continuum transport model based on pore-scale characteristics. Our approach relies on building a simplified unit cell made up of immobile and mobile continua. We employ a numerically resolved pore-scale velocity distribution to characterize the volume of each continuum and to define the velocity profile in the mobile continuum. Using the simplified unit cell, we derive a closed form model, which includes two effective parameters that need to be estimated: a characteristic length scale and a ratio of waiting times $R_D$ that lumps the effect of stagnant regions and escape process. To calibrate and validate our model, we rely on a set of pore-scale numerical simulation performed on a 2D disordered segregated periodic porous medium considering different initial solute distributions. Using a Global Sensitivity Analysis, we explore the impact of the two effective parameters on solute concentration profiles and thereby define a sensitivity analysis driven criterion for model calibration. The latter is compared to a classical calibration approach. Our results show that, depending on the initial condition, the mass exchange process between mobile and immobile continua impact on solute profile shape significantly. By introducing parameter $R_D$ we obtain a flexible transport model capable of interpreting both symmetric and highly skewed solute concentration profiles. We show that the effectiveness of the calibration of the two parameters closely depends on the content of information of calibration dataset and the selected objective function whose definition can be supported by of the implementation of model sensitivity analysis. By relying on a sensitivity analysis driven calibration, we are able to provide a good interpretation of the concentration profile evolution independent of the given initial condition relying on a unique set of effective parameter values.

# 1 Introduction

The development of accurate mathematical models to describe solute mass transport in porous media is particularly challenging when the medium is characterized by the presence of cavities, dead-end pores, stagnant zones and a highly heterogeneous velocity field. The structure and the extent of low-velocity regions directly impacts solute transport, potentially leading to long mass retention times. Accurately modeling such trapping effects is crucial, for example, in the context of remediation and risk assessment (e.g. *de Barros et al.,* 2013). A sound understanding of the conditions and pore-scale processes that physically control the rate of exchange between stagnant and fast-flowing regions is needed to better understand solute spreading and mixing and subsequently the evolution of conservative and reactive transport processes (e.g. *Alhashmi et al.*, 2015; *Lichtner and Kang*, 2007; *Kitanidis and Dykaar*, 1997; *Wirner et al.*, 2014; *Cortis and Berkowitz*, 2004; *Briggs et al.*, 2018; *Baveye et al.*, 2018).

An appealing approach to study solute transport is to perform direct numerical simulations at pore scales (see e.g. *Scheibe et al.,* 2015; *Bijeljic et al.*, 2013; *Hochstetler et al.*, 2013; *Porta et al.*, 2013). Such techniques present the remarkable advantage of providing detailed information on solute concentration evolution at each point of the porous domain. However, the applicability of such methods, which are computationally demanding, is limited to small domains, typically much smaller than field scales of common interest (*Dentz et al., 2011*). Upscaled continuum models are consequently more suited to simulating larger-scale systems. In particular, double or multiple continua approaches are appealing due to their ability to explicitly distinguish stagnant zones from fast flowing channels.

In the classical double-continuum approach (*Coats and Smith*, 1964; *Haggerty and Gorelick*, 1995; *Carrera et al.*, 1998; *Bear and Cheng*, 2010), a mobile and immobile continuum exchange mass as a first order process with an effective mass transfer coefficient. Typically, effective parameters of double or multi-continuum models, need to be estimated via fitting against solute concentration data, e.g., measured breakthrough curves or solute concentration profiles.

Alternatively, up-scaled dual continua models have been formally derived (see e.g., *Souadnia et al.*, 2002) by means of volume-averaging techniques. While appealing due to their sound theoretical basis, as discussed by *Davit et al.* (2012), these formally derived double-continua formulations present practical limitations in terms of their applicability to real problems. Often such models can be nearly as difficult to solve as their pore-scale direct simulation counterparts, as they include highly complex non-local terms, which imply significant numerical implementation challenges (*Porta et al.*, 2016).

To overcome such limitations, a number of studies have proposed less formal, but more parsimonious effective upscaled formulations that still exploit key pore scale information. These studies encompass Eulerian (*Porta et al.*, 2015) and Lagrangian formulations (*Sund et al.*, 2017a,b; *Dentz et al.*, 2018). These methods are designed to embed pore-scale characteristics into parameters which can be applied at a larger scale. For example, *Sund et al.* (2017a,b) employ trajectories and travel times distributions measured numerically at pore-scale to infer the effective evolution of mixing and reaction rates in an effective Lagrangian spatial Markov model. The work of *Porta et al.* (2015) focuses on the use of pore-scale information to characterize a double-continuum transport model, from an Eulerian perspective. *Porta et al.*'s (2015) model relies on the cumulative distribution function of velocity measured from a pore-scale simulation of single-phase flow and assumes that the exchange time between high and low velocity regions is dictated by the characteristic diffusion time scale. The model reproduces observed transport behaviors in relatively well connected three-dimensional porous systems, i.e. a beadpack and a sandstone sample. However, as shown by *Bénichou and Voituriez* (2008), realistic cavities may be characterized by complex geometry such that it can take an extremely long time to exchange mass from slow regions to faster flowing channels. For this reason, the double-continuum approach proposed by

*Porta et al.* (2015) might suffer limitations when the geometry of the porous medium is highly tortuous and presents significant stagnant cavities (e.g. carbonate rocks, see *Bijeljic et al.*, 2013).

In order to overcome these limitations, in the framework defined by *Porta et al.* (2015), we develop a double-continuum model, which explicitly accounts for a characteristic time for the exchange process between high and low velocity regions which may be larger than the diffusion time scale. This model lumps the effect of exchange process at pore-scale into a single effective parameter, which is defined as the ratio of the time required by the solute to escape/explore the stagnant regions of the porous medium to the characteristic diffusive time scale. Our main objective is to derive a closed form double continuum model and to test it against numerical pore-scale simulations of solute transport performed in a disordered synthetic two-dimensional porous medium, considering different initial conditions.

We explore the flexibility of the model by means of a sensitivity analysis. We assess the effectiveness of the model by means of *i*) a qualitative inspection of concentration profiles predicted considering different initial conditions and *ii*) quantification of the Sobol' indices of appropriately defined target metrics. We also investigate the role of a global sensitivity analysis (GSA) in defining a tailor-made objective function to increase the efficacy of model calibration.

The paper is structured as follows. In Section 2 we present the problem setup that will be used as the test bed for the proposed double continuum model, along with details on the pore-scale model. In Section 3, we derive the proposed closed form double continuum model. In Section 4, the flexibility of model is explored via a GSA. Calibration and validation of the model are discussed in Section 5 and conclusions are presented in Section 6.

# 2 Problem Setting

## 2.1 Pore scale domain

In this work we consider a two-dimensional porous medium made up of repeating periodic unit cells, $\Omega'$. The cell configuration is the same as that of *Porta et al.* (2016). The geometry of the unit cell is generated by the disordered superposition of circular grains of uniform diameter $w = 8 \times 10^{-5}$ m and is characterized by porosity $\phi = 0.5948$. The domain is periodic in both longitudinal and transversal directions. The geometry of the unit cell and the associated normalized velocity field (normalized such that the mean velocity modulus is equal to 1) is shown in Figure 1a. The velocity field is computed by solving the incompressible Navier-Stokes equations, assuming a fully saturated medium (see *Porta et al.*, 2016; *Bekri et al.*,1995; *Coelho et al.*, 1997). The velocity field is resolved on a regular spatial grid with a resolution of $2 \times 10^{-5}$ m. A unit cell has total dimension of $4.8 \times 10^{-3}$ m (longitudinal to main flow) times $1.2 \times 10^{-3}$ m (transverse). The total length $D$ of the porous domain is equal to 0.192 m, corresponding to a sequence of 40 unit cells. In dimensional terms the average velocity along the $x$-axis ($U$ [m s$^{-1}$]) is equal to $6.22 \times 10^{-5}$ m s$^{-1}$. The velocity field shows large variability, ranging over 10 orders of magnitude. We define low-velocity regions as those below $10^{-2}$ in the normalized velocity field, which are segregated from the well-connected higher-velocity channels. We use this value to distinguish between the disconnected Low-Velocity (LV) regions and the continuous High-Velocity (HV) channels. In Figure 1b we explicitly identify the solid phase and the LV and HV regions. Based on our chosen threshold, we can split the total porosity $\phi$ into two parts: $\phi_{HV} = 0.5131$ and $\phi_{LV} = 0.0817$ associated with the HV and LV regions, respectively.

## 2.2 Initial conditions

As a test bed for the double-continuum model that we will present in Section 3, we perform a series of pore-scale simulations of conservative transport with three different initial scenarios, labeled S_U, S_HV and S_LV:

- S_U: in one unit cell the solute is uniformly placed in both the HV and LV regions. We define the initial concentration as $\overline{E}_0$.
- S_HV: solute is uniformly placed in only the HV region for the extent of one unit cell. The cross-sectional averaged initial concentration in the HV region ($\overline{E}_{0HV}$) is such that $\overline{E}_0 = \overline{E}_{0HV}\, \phi_{HV} / \phi$.
- S_LV: solute is uniformly placed in the LV region for the extent of one unit cell. The cross-sectional averaged initial concentration in the LV region ($\overline{E}_{0LV}$) is such that $\overline{E}_0 = \overline{E}_{0LV}\, \phi_{LV} / \phi$.

The aforementioned initial conditions are chosen to mimic conditions of particular interest when considering, for example, a contaminated site (e.g. *de Barros et al.*, 2013) or for the interpretation of experimental results, where solute injection in a column is typically flux weighted while small scale propagators allow observing the displacement of a resident solute.

2.3 Pore scale numerical simulations

Pore-scale modeling of the concentration field is performed in a particle tracking framework, with the time domain random walk (TDRW) approach described in detail in *Russian et al.* (2016). This approach is particularly suitable for simulations in media that display a broad range of velocity such at the one considered here (*Banton et al.*, 1997; *Delay et al.*, 2002, *Bodin*, 2015). Its benefit is that particles do not spend a large number of random walk steps in low velocity regions due to a constant time discretization, typical of classical particle tracking methods. The TDRW method is formally equivalent to a discretized advection-diffusion equation (*Russian et al.*, 2016). At the same time, its formulation coincides with a continuous time domain random walk (CTRW) with space-dependent transition times and probabilities.

Here we use 2 million particles for each simulation. For each initial condition illustrated in Section 2.2, we explore the temporal evolution of concentration over a total time of 400 s. Profile concentrations are obtained by vertical integration of particle numbers normalized by the corresponding porosity of the vertical slice.

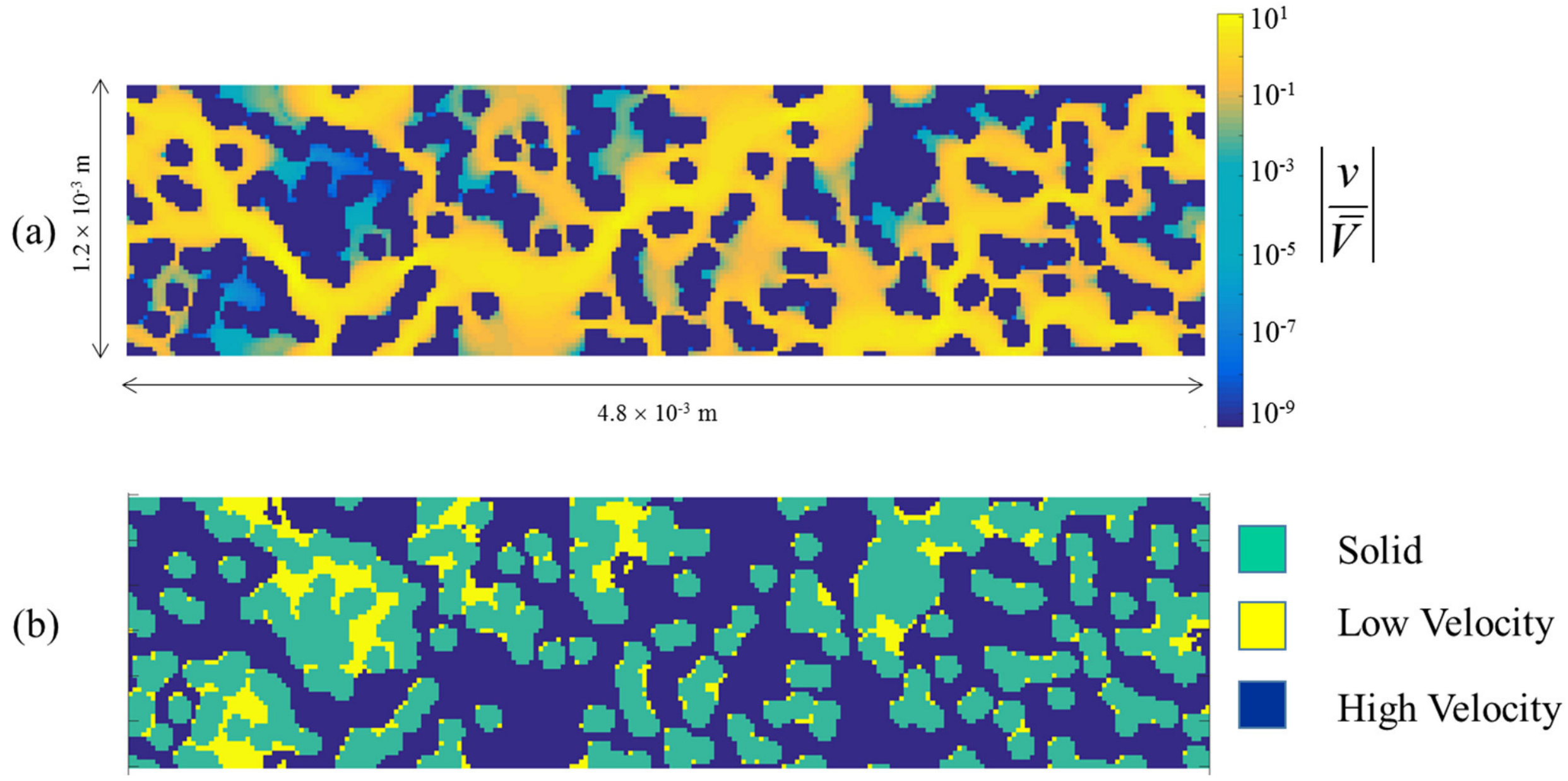

Figure 1: (a) the geometry of the unit cell with the associated normalized velocity field and (b) the geometry of the low velocity (yellow) and high veleocity (blue) regions resulting from imposing $10^{-2}$ as normalized velocity threshold to discriminate between the low and high velocity regions.

## 3 Dual continuum model formulation

The development of the double-porosity model proposed here is built starting from the procedure originally developed by *Porta et al.* (2015) and it is schematically outlined in Figure 2. We start with the 2D-porous medium introduced in Section 2 (Figure 1). We define the average Péclet number $Pe_{av} = UL/D_m$ where $D_m$ [m$^2$ s$^{-1}$] is the molecular diffusion coefficient and $L$ [m] a characteristic length scale of the system that is considered unknown a priori and should be properly determined.

In the double-continuum model the porous system is then conceptualized as a simplified unit cell of thickness $L$ with a uniform shear flow (Figure 2a). The direction of flow is only in the $\hat{x}$ [m] direction.

The unit cell is split in two parts, labeled mobile and immobile. The mobile zone (light color in Figure 2a) occupies $-l/2 < \hat{y} < l/2$ (with $\hat{y}$ [m]) and the immobile zone (dark color in Figure 2a) $-L/2 < \hat{y} < -l/2$ and $l/2 < \hat{y} < L/2$. The value of $l$ [m] is computed such that $l/L = \phi_{HV}/\phi$ and $(L-l)/L = \phi_{LV}/\phi$ to partition the mobile/immobile region in the elementary cell in the same manner as the HV and LV regions in the reference porous medium. The velocity distribution in the mobile zone is the (appropriately rescaled) sample cumulative distribution function (*cdf*) from pore-scale velocities belonging to the HV region of the porous medium. In Figure 2b, we depict the *cdf* computed for the normalized fluid velocity. The vertical dashed line represents the chosen threshold discriminating between HV and LV regions.

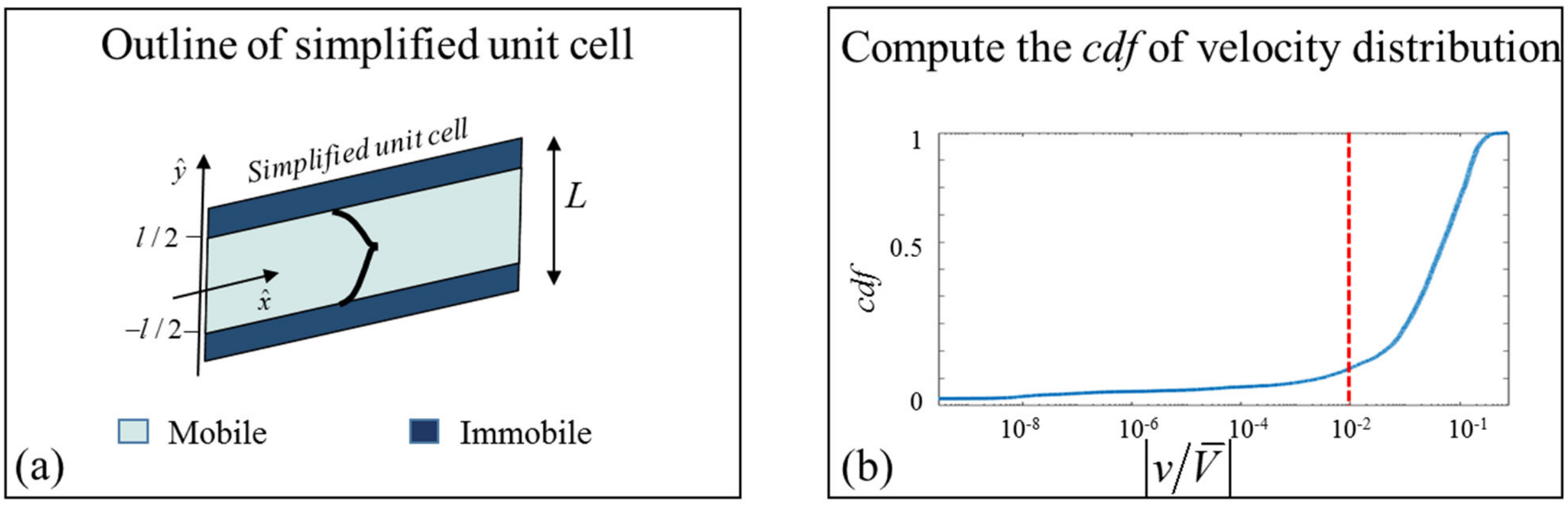

Figure 2: (a) outline of the simplified unit cell and (b) definition of the Cumulative Density Function (*cdf*) of the local velocities of the fluid phase.

We assume that solute transport in the unit cell is described by the following dimensionless system of equations

$$\begin{cases} \dfrac{\partial E_M}{\partial t} + u\dfrac{\partial E_M}{\partial x} = \dfrac{1}{Pe}\dfrac{1}{\tau_M}\dfrac{\partial^2 E_M}{\partial x^2} + \dfrac{1}{Pe}\dfrac{\partial^2 E_M}{\partial y^2} & |y| < l/(2L) \\ \dfrac{\partial E_I}{\partial t} = \dfrac{1}{Pe}\dfrac{1}{\tau_{IM}}\dfrac{\partial^2 E_I}{\partial x^2} + \dfrac{R_D}{Pe}\dfrac{\partial^2 E_I}{\partial y^2} & |y| > l/(2L) \end{cases} \tag{1}$$

completed by the following boundary conditions

$$E_M = E_I \text{ and } \frac{\partial E_M}{\partial y} = R_D \frac{\partial E_I}{\partial y} \qquad |y| = l/(2L) \qquad \text{(a)}$$
$$\frac{\partial E_I}{\partial y} = 0 \qquad |y| = l/(2L) \qquad \text{(b)} \qquad (2)$$

where $E_M(x,y,t) = C_{EM}(x,y,t)/c_0$ [-] and $E_I(x,y,t) = C_{EI}(x,y,t)/c_0$ [-] are dimensionless concentrations in the mobile (HV) and immobile (LV) zones of solute *E*, respectively; $C_{EM}(x,y,t)$ [mol L$^{-1}$] and $C_{EI}(x,y,t)$ [mol L$^{-1}$] are the concentrations in mobile and immobile region ([mol L$^{-1}$]) and $c_0$ ([mol L$^{-1}$]) is a characteristic concentration; $u$ [-] represents the dimensionless velocity along the *x*-direction computed as the ratio between the dimensional velocity $\hat{u}$ (m s$^{-1}$) and the mean velocity in the mobile region $U_M$ [m s$^{-1}$], i.e. $u = \hat{u}/U_M$ ; *t* [-] is the dimensionless time defined as $t = \hat{t}/t_a$ where $\hat{t}$ [s] is the time and $t_a = L/U_M$ [s] is the advective time scale; $x = \hat{x}/L$ [-] and $y = \hat{y}/L$ [-] are the reference system coordinates; *Pe* [-] is the Péclet number computed based on the mean mobile velocity as $Pe = U_M L / D_m$ .

We introduce in (1) two key novel elements with respect to the formulation of *Porta et al.* (2015):

- We include the dimensionless parameters $\tau_M$ and $\tau_{IM}$ that represent the tortuosity factors associated with the HV and LV regions respectively. Including $\tau_M$ and $\tau_{IM}$ embeds the impact of the phase geometry on the evolution of solute diffusion along the *x*-direction in the simplified unit cell (*Shen and Chen*, 2007). The values of $\tau_M$ and $\tau_{IM}$ are computed directly using the HV and LV geometries, by mean of the TauFactor Matlab code (*Cooper et al., 2016,* MATLAB® and Statistics Toolbox Release 2016b). The HV region is characterized by $\tau_M = 2.48$ while $\tau_{IM}$ is infinite since the LV region is clearly disconnected (Figure 1b). The latter is a consequence of considering a two-dimensional porous medium. In more realistic 3D systems, the LV region is connected due to the no-slip boundary condition imposed at solid-fluid interface and the fact that solid phases connect (unlike in 2D). Here, to mimic a disconnected LV region, we imposed a very high value of tortuosity, $\tau_{IM} = 100$.

- we introduce parameter $R_D$ in (2.a), defined as

$$R_D = \frac{t_D}{t_e} \tag{3}$$

where $t_e$ [s] is the characteristic time required by the solute to escape/explore the stagnant regions of the porous medium while the $t_D$ [s] is the characteristic time scale of the diffusion process in a free fluid.

In *Porta et al.* (2015), mass exchange between HV and LV regions is modeled as a diffusive process at the interface of immobile and mobile zones, which means that the characteristic time of the mass exchange is assumed equal to the characteristic diffusion time $t_D = L^2 / D_m$. This choice may not always be appropriate, for example if the porous structure includes large stagnant cavities connected to fast channels through narrow bottlenecks. Indeed, pore-scale and theoretical investigations (e.g. *Wirner et al.*, 2014; *Bénichou and Voituriez*, 2008; *van Genuchten*, 1985; *Holcman and Schuss*, 2013) show that solute mass enclosed in a stagnant (or fast-flowing) region may take an extremely long time to escape (or explore) the zone depending on its shape.

Applying the same averaging procedure implemented by *Porta et al.* (2015) on the system (1)-(2), the closed section averaged form of the proposed model reads (see the Supplementary Material for more details)

$$\begin{cases} \dfrac{\partial \overline{E}_M}{\partial t} + \dfrac{\partial \overline{E}_M}{\partial x} + \dfrac{\partial}{\partial x}\left[ d_{H1} \dfrac{\partial \overline{E}_M}{\partial x} + d_{H2} \Delta \overline{E}_{MI} \right] = \dfrac{1}{Pe} \dfrac{1}{\tau_M} \dfrac{\partial^2 \overline{E}_M}{\partial x^2} + \dfrac{\phi}{Pe \phi_{HV}} \left( e_1 \dfrac{\partial \overline{E}_M}{\partial x} + e_2 \Delta \overline{E}_{MI} \right) \\ \dfrac{\partial \overline{E}_I}{\partial t} = \dfrac{1}{\tau_{IM}} \dfrac{1}{Pe} \dfrac{\partial^2 \overline{E}_I}{\partial x^2} - \dfrac{\phi}{\phi_{LV} Pe} \left( e_1 \dfrac{\partial \overline{E}_M}{\partial x} + e_2 \Delta \overline{E}_{MI} \right) \end{cases} \tag{4}$$

where $\bar{E}_M(x,t)$ and $\bar{E}_{IM}(x,t)$ are the averaged (along the *y*-direction in the cell) concentrations of the solute in the mobile and immobile regions, respectively while $\Delta\bar{E}_{MI} = \bar{E}_M - \bar{E}_{IM}$. The effective parameters $d_{H1}$, $d_{H2}$, $e_1$ and $e_2$ appearing in (4) are defined as follow

$$d_{H1} = \frac{L}{l}\int_{-l/(2L)}^{l/(2L)} b_1\tilde{u}\,\mathrm{d}y; \quad d_{H2} = \frac{L}{l}\int_{-l/(2L)}^{l/(2L)} b_3\tilde{u}\,\mathrm{d}y \tag{5}$$

$$e_1 = 2R_D\left.\frac{\partial b_2}{\partial y}\right|_{y=l/(2L)} = 2\left.\frac{\partial b_1}{\partial y}\right|_{y=l/(2L)}; \quad e_2 = 2R_D\left.\frac{\partial b_4}{\partial y}\right|_{y=l/(2L)} = 2\left.\frac{\partial b_3}{\partial y}\right|_{y=l/(2L)} \tag{6}$$

Here $b_i$ ($i$=1,2,3,4) are four closure variables and $\tilde{u}$ indicates the fluctuation of the velocity along the $x$ direction with respect to the mean velocity. The set of closure variables $b_i$ ($i$=1,..,4) can be solved numerically given the known $u(y)$ and $b_i(y)$ in the simplified unit cell. Details on the closure problem and its solution are provided in the Supplementary Material.

Investigating the specific relationship between porous geometry structure and $R_D$ is beyond the scope of the present work and is postponed to future efforts, but understanding its role on large-scale transport is considered next. For the purposes of this work, the quantity $R_D$ is considered an effective parameter along with the characteristic length scale $L$ whose characterization based on pore-scale geometry feature is still open to debate and different methodologies have been proposed in literature (see e.g. *Bear and Cheng*, 2010; *Mostaghimi et al.*, 2012; *Siena et al.* 2014; *Mayer and Bijeljic*, 2016; *Alhashmi et al.*, 2016; *Dullien*, 2012).

# 4 Characterization of the mass transfer at continuum scale

In this section we elucidate how the proposed model accounts for exchange process between fast and slow regions and the impact of the exchange process on solute evolution depending on initial conditions.

## 4.1 Quantification of mass transfer time

As proxy to quantify the exchange process simulated by the model we introduce

$$Q(\hat{t}) = \frac{\left|\langle E_I(\hat{t})\rangle - \langle E_M(\hat{t})\rangle\right|}{E_0} \tag{7}$$

where $\hat{t}$ [s] is time, $E_0$ is initial concentration of the conservative solute, $\langle E_I(\hat{t})\rangle$ and $\langle E_M(\hat{t})\rangle$ are the average concentration in the immobile and mobile zones along the $x$-direction, $D$ respectively, i.e.,

$$\langle E_j\rangle = \frac{\int_D \bar{E}_j dx}{V_j} \qquad \text{with } j\text{=I, M} \tag{8}$$

with $V_M$ and $V_I$ equal to the pore volume associated with the HV and LV regions of the porous domain. An illustrative example of $Q(\hat{t})$ computed for initial condition S_LV and four combinations of $L$ and $R_D$ is shown in Figure 3a. $Q(\hat{t})$ decreases monotonically from 1 to 0 over time. When $Q(\hat{t})$ is equal to 1 all of the solute mass is in the immobile zone, while $Q(\hat{t})$=0 indicates an equilibrium between the solute mass flux escaping and entering the immobile zone. How quickly $Q$ reaches zero tells us how quickly the exchange process occurs. The red line ($L$=1000µm; $R_D$= 1 × $10^{-4.72}$) and blue line ($L$=98µm; $R_D$= 1 × $10^{-0.16}$) in Figure 3a illustrate two extremes cases. For large $L$ and small $R_D$, $Q(\hat{t})$ suggests an extremely slow escape process such that at 400 s the solute is still almost totally trapped in the immobile zone. This is confirmed in Figure 3b where we display solute concentration profiles at $\hat{t}$ =200s for the same combination of $L$ and $R_D$ values of in Figure 3a. The yellow area identifies the initial step condition. The concentration profile for $L$=1000µm and $R_D$= 1 × $10^{-4.72}$ (red line) coincides closely with the initial condition. By contrast, for small $L$ and $R_D$ close to one, the exchange process is virtually instantaneous since $Q(\hat{t})$ approaches zero in the first time step of the simulation (see blue line in Figure 3a). The corresponding concentration profile (blue line in Figure 3b) does not present heavy tailing nor strong asymmetry suggesting that the retentive effect of the stagnant regions is negligible. The black and magenta lines represent intermediate conditions demonstrating that both $L$ and $R_D$ influence the mass exchange process. The four different profiles in Figure 3b present completely different shapes and spreading patterns, demonstrating that within the context of the proposed model accessibility of the stagnant regions can significantly impact the profile evolution, even when starting from the same initial condition (i.e., S_LV).

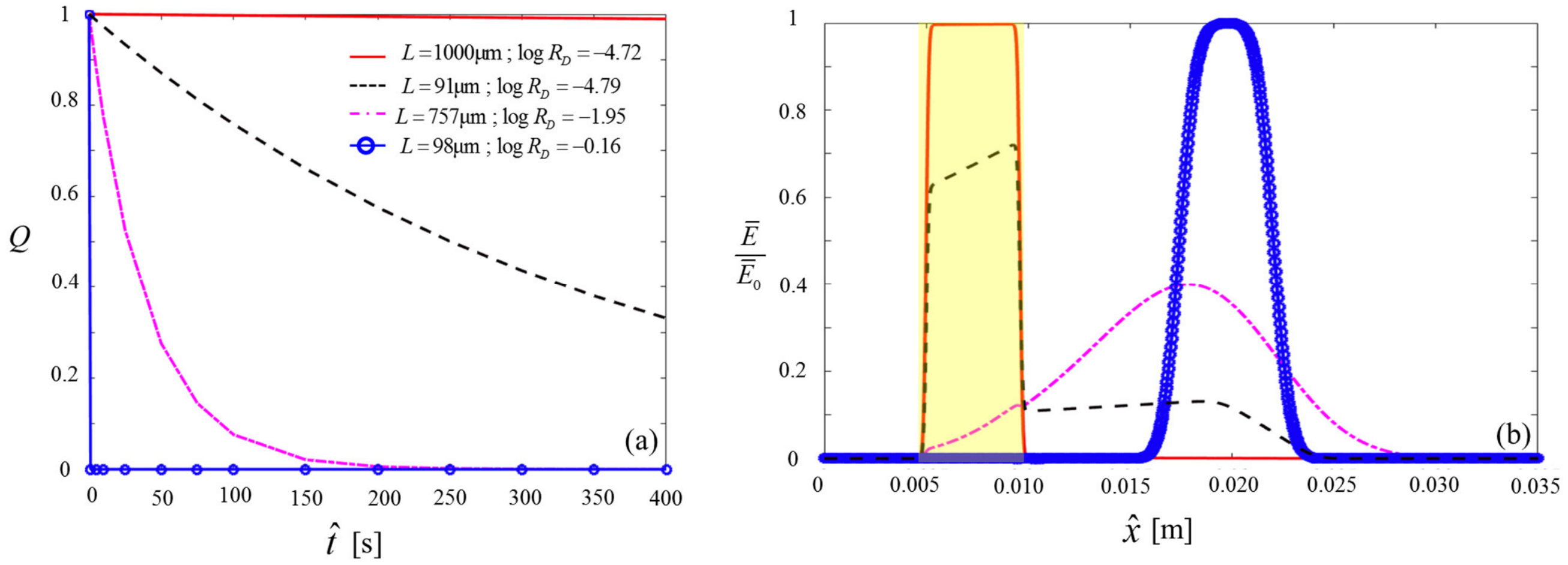

Figure 3: (a) Temporal evolution of $Q(t)$ for four different combinations of the effective parameters $L$ and $R_D$ computed relying on the S_LV initial scenario; (b) Concentration profiles yielded by model implementation at $\hat{t}$ = 200 s for the same four combinations of parameter values and initial scenario reported in panel (a). The shadowed yellow area identifies the location of the solute in the initial condition.

## 4.2 Impact of initial condition

The value of effective parameters embedded in formulation (4)-(6) should depend solely on the porous medium structure and geometry of cavities. This implies that the parameter values should not depend on the initial condition being investigated. By considering different initial conditions as introduced in Section 2, we can compare concentration profiles at $\hat{t}$ =200s yielded by fixing the values of effective parameters $L$ and $R_D$ (Figure 4). The combinations of parameters selected for Figure 4 are those in Figure 3 for the S_LV scenario. When the exchange process is very fast (magenta and blue lines in Figure 3a), the impact of the initial condition is virtually not detectable (Figure 4d). As the exchange process slows down, the initial condition impacts predictions more markedly. S_HV and S_U lead to similar profiles (showing similar spread and skewness); the only notable difference is that the solute with S_U is prone to develop a thicker backward tail (see Figure 4a, 4b and 4c) as part of the solute is initially entrapped in the immobile zone. The profiles generated starting from the S_HV scenario present a symmetric shape for both very fast and extremely slow exchange process (see Figure 4b and d). The solute profiles features associated with S_LV are markedly different, trailing the other cases as solute has to leave the trapped phase before beginning its downstream journey.

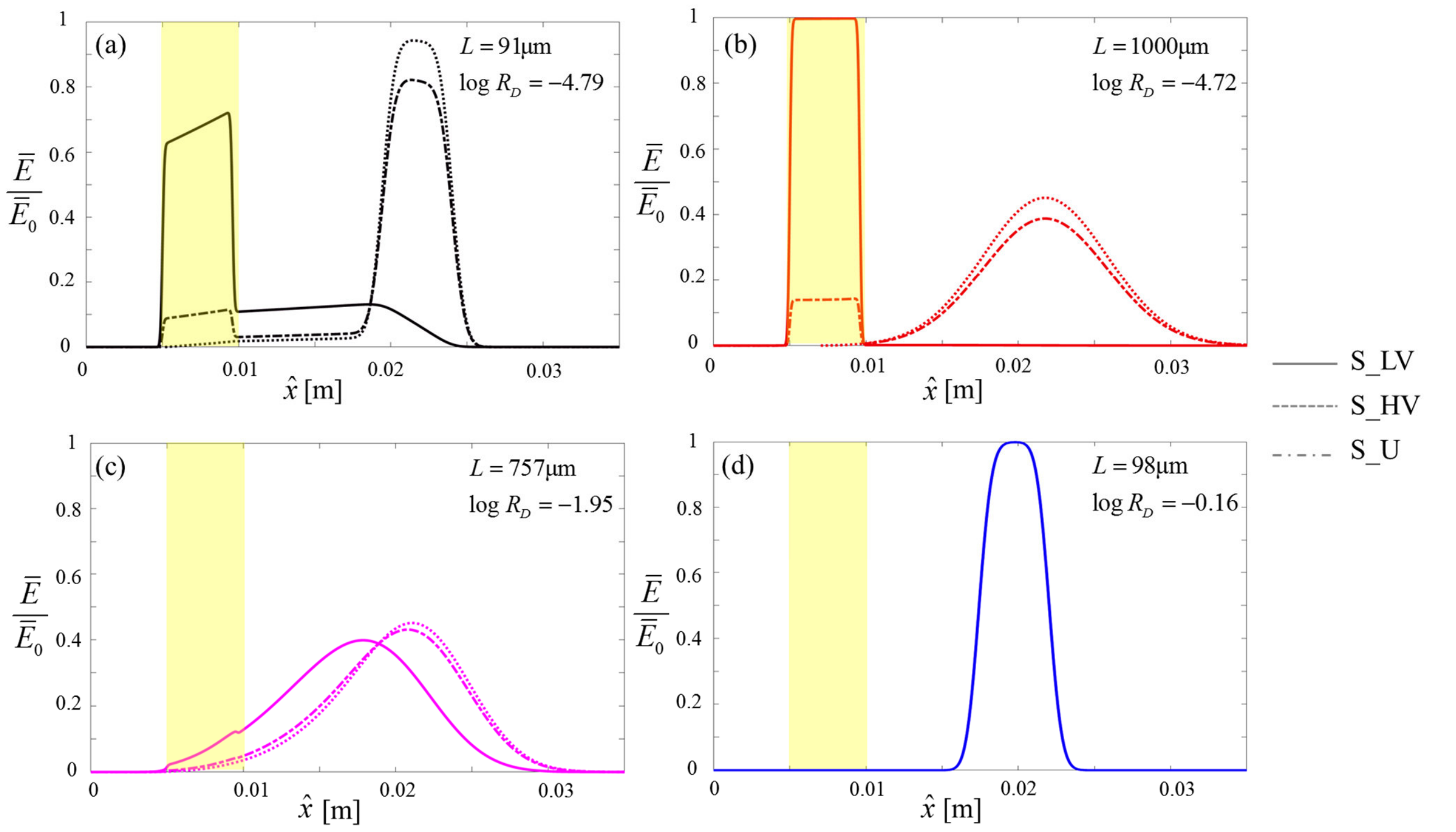

Figure 4: Concentration profiles given by solution of equations (4)-(6) at $\hat{t}$ = 200 s for S_U, S_HV and S_LV initial conditions with (a) $L$ = 91 μm and $R_D = 10^{-4.79}$, (b) $L$ = 1000 μm and $R_D = 10^{-4.72}$, (c) $L$ = 757 μm and $R_D = 10^{-1.95}$ and (d) $L$ = 98 μm and $R_D = 10^{-0.16}$. The shadowed yellow areas indicate the location where the solute is initially placed.

## 4.3 Global Sensitivity Analysis

The qualitative analysis illustrated above provides a quick understanding of the impact of the exchange process on model predictions, but is limited to only four combinations of effective parameters selected for presentation purposes. To explore more extensively and rigorously the impact of the variability of the effective parameters on solute transport depending on the initial condition, we investigate different model responses by means of a Global Sensitivity Analysis (GSA; *Saltelli et al.*, 2008; *Sobol*, 1993). Our goal is to explore the sensitivity of the mass transfer process to $L$ and $R_D$ and thereby the resulting impact of mass transfer on the shape of longitudinal concentration profiles under different scenarios. To this end we define the following outputs

$$T_{50} = \{\hat{t} \mid Q(\hat{t}) = 0.5\} \tag{9}$$

$$\sigma^2(\hat{t}) = \int_D \left[\hat{x} - \mu(\hat{t})\right]^2 \bar{E}_n(\hat{x},\hat{t})d\hat{x} \tag{10}$$

$$\gamma(\hat{t}) = \frac{\int_D \left[\hat{x} - \mu(\hat{t})\right]^3 \bar{E}_n(\hat{x},\hat{t})d\hat{x}}{\sigma^3(\hat{t})} \tag{11}$$

where $\bar{E}_n$ is the section-averaged concentration normalized to total solute mass present in the system and $\mu(\hat{t})$ is the first spatial moment of $\bar{E}_n$. Here, $T_{50}$ [s] is a characteristic time for mass transfer, while $\sigma^2$ and $\gamma$ quantify spreading and skewness of the solute concentration profile. We assume the two effective parameters embedded in the model (i.e. $R_D$ and $L$) to be two independent random uniformly distributed variables. The parameter $L$ is assumed to vary in $\Omega_L$ defined between 80 μm, the diameter of the cylinders used to generate the porous medium (see Section 2), and 1200 μm, i.e. the total length of the periodic unit cell in the transverse direction. The parameter space of $R_D$ ($\Omega_{RD}$) is bounded between 1 and $10^{-5}$. The upper bound corresponds to a porous structure that is easily accessible by the solute with the escape/exploration time of the cavities equal to $t_D$. The lower bound is estimated based on the results of *Wirner et al.* (2014), who investigate trapping effects of stagnant zones in a quasi two-dimensional porous medium, similar in nature to the one we use here. Using $N$ quasi-Monte Carlo (Sobol, 1998) samples ($N$=1000) of the parameter space $\Omega_L \times \Omega_{RD}$, we run the model for each of the three different initial conditions and compute the first order Sobol' indices associated with the three target variables (9) -(11).

Note that as we consider only two parameters, we can state the following equality by the ANOVA (ANalysis Of VAriance, *Sobol*, 1993) decomposition of variance

$$SI(k)_L + SI(k)_{RD} + SI(k)_{L,RD} = 1 \text{ with } k = T_{50}, \sigma^2, \gamma \tag{12}$$

where $SI(k)_L$ and $SI(k)_{RD}$ are the first order Sobol' indices of variable $k$ associated with parameters $L$ and $R_D$, respectively, while $SI(k)_{L,RD}$ is the second order Sobol' index quantifying two parameter interaction effects on target quantity $k$.

Figure 5 shows that the characteristic time associated with mass transfer $T_{50}$ is influenced by both parameters, so that its variability is explained mostly by the combined variation of the two parameters. This result is quantified by Sobol' indices, which show the largest contribution to the variance of $T_{50}$ is given by the combined effect of the two effective parameters ($SI(T_{50})_{L,RD}=0.55$, see Table 1). This essentially means that information on both parameters ($L$ and $R_D$) can help better constrain this output (and vice versa, i.e. this output, if measurable, could be used to constrain both parameters).

The signature of the results on $T_{50}$ is reflected in a different fashion on profile shapes for different scenarios, as shown in Figure 6. The Sobol' indices indicate that for scenario S_LV spreading is mainly influenced by the combined variability of $L$ and $R_D$, because $SI(\sigma^2)_L + SI(\sigma^2)_{RD} < 0.4$ for all the considered times. This is explained by the fact that for S_LV solute displacement is limited by mass transfer to high velocity regions from low velocity regions, where the solute is initially residing. A similar result is obtained for the skewness $\gamma$ limited to early times. As time progresses, $\gamma$ appears to be chiefly influenced by $R_D$ for S_LV, while the influence of $L$ becomes relatively small after $\hat{t}$ = 50 s. In the S_HV scenario, parameter $L$ almost entirely controls the variance of $\sigma^2$ and the chosen length scale has a prominent role on solute spread in the mobile region. In this scenario, the solute accesses low velocity regions by mass transfer while being transported downstream. Parameter $L$ has an effect on the $Pe$ number and thereby on the spreading of the profile due to dispersion in the mobile region (i.e., dispersive parameters in equation (4) increase with $L$). The effect of $R_D$ and the coupled effect between the two parameters emerges more clearly when profile skewness $\gamma$ is considered. In particular, $\gamma$ is chiefly influenced by the joint variation of the two parameters ($SI(\gamma)_L + SI(\gamma)_{RD} < 0.6$) for $\hat{t} > 50$ s. These results show that solute mass transfer has a negligible influence on spreading for S_HV, but is persistently reflected in the profile skewness at long times.

In scenario S_U mass distribution between high and low velocity distributions is at equilibrium in the initial condition, i.e. the effect of mass transfer can be expected to be minor compared with the other two cases. Consistent with this observation, spreading $\sigma^2$ is initially controlled exclusively by $L$, while with time the situation is gradually reversed, i.e. $R_D$ becomes the most important parameter. At all times $SI(\sigma^2)_L + SI(\sigma^2)_{RD} \approx 1$ and $SI(\gamma)_L + SI(\gamma)_{RD} > 0.8$, showing that the two parameters act independently to influence the profile shape in scenario S_U.

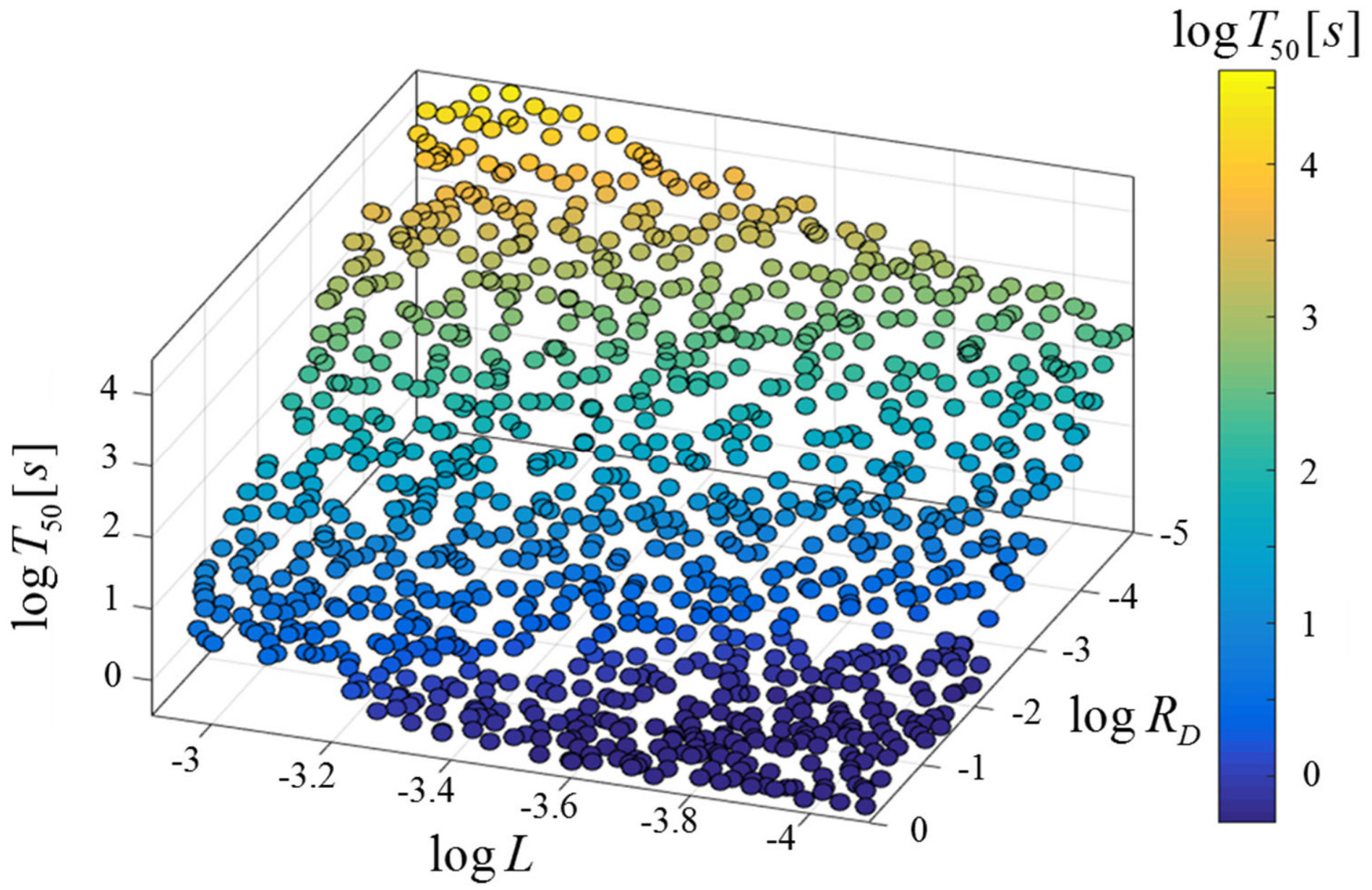

Figure 5: Evolution of the log $T_{50}$ across the parameter space ( i.e., log $L$ and log $R_D$ ) for the $N$ sampled combinations of $L$ and $R_D$.

Table 1: First and second order Sobol' indices computed for the target variable $T_{50}$.

| Index | Sobol' indices value |
|---|---|
| $SI(T_{50})_L$ | 0.12 |
| $SI(T_{50})_{RD}$ | 0.33 |
| $SI(T_{50})_{L,RD}$ | 0.55 |

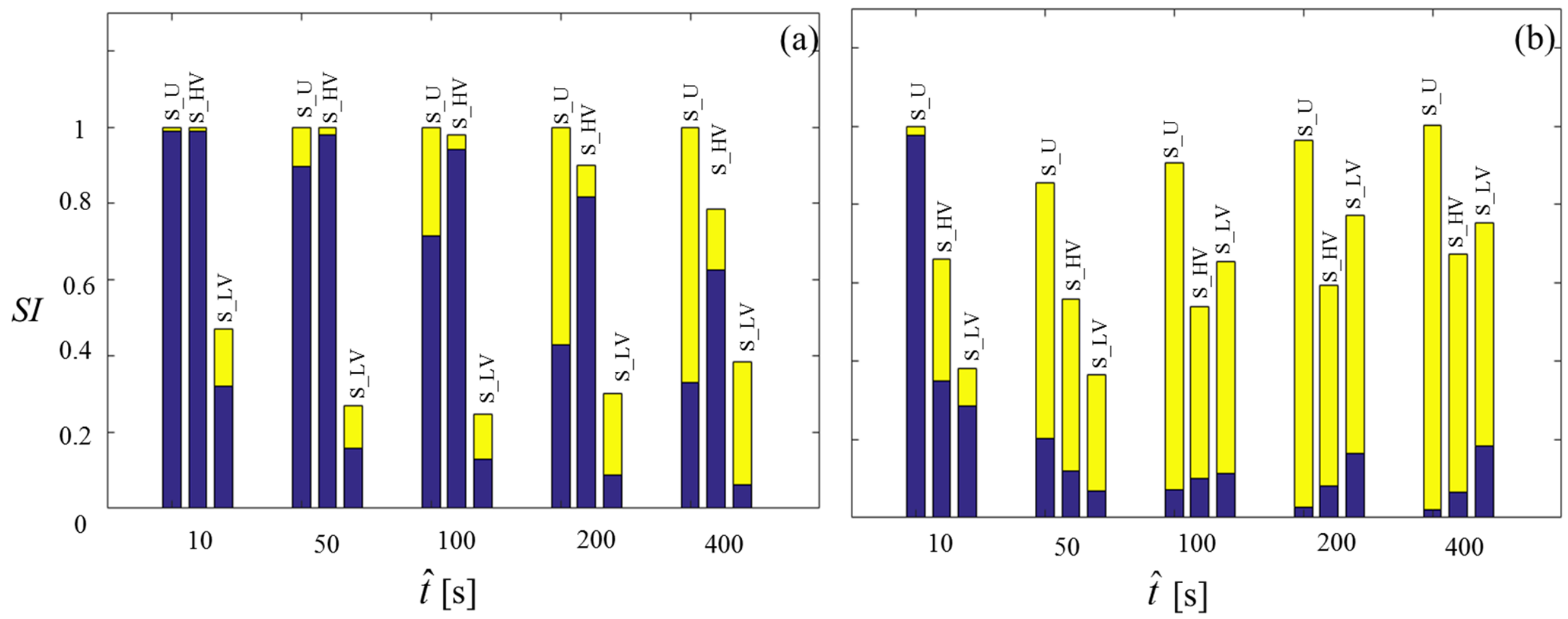

Figure 6: Sobol' indices computed for (a) the spread of the concentration profile ($\sigma^2$) and (b) the skewness of the concentration profiles ($\gamma$). Both panels (a) and (b) report the Sobol' indices at five time levels ($\hat{t}$ = 10, 50, 100, 200 and 400 s) for the three different initial scenarios investigated in this work (S_U, S_HV and S_LV). The blue and yellow portions of the bars quantify $SI(k)_L$ and $SI(k)_{RD}$, respectively, with $k=\sigma^2$, $\gamma$.

# 5 Model Calibration and Validation

In Section 4, we showed that the proposed model is flexible and able to reproduce symmetric profiles, highly skewed profiles or entrapped solute for extremely long time by opportunely setting two effective parameters $L$ and $R_D$. In this Section, we discuss calibration and validation of the model against pore-scale simulations performed in the two-dimensional porous medium from Section 2.

## 5.1 Model calibration

Calibration of the continuum model is performed thorough the minimization of two different objective functions: *OBF* and a Sensitivity Oriented Objective Function (*SOOF*). We define *OBF* as

$$OBF = \sum_{i=1}^{DC} \left[ E(\hat{x}_i) - E_i^* \right]^2 \tag{13}$$

where $\bar{E}(\hat{x}_i)$ is the concentration from the continuum model at $\hat{x}_i$ while $E_i^*(\hat{x}_i)$ is the section average concentration from the pore-scale simulations. *DC* is the number of data used in the calibration. To compute $E^*$, we use the same discretization as for the velocity field grid. Objective functions like that in (13) are commonly used to estimate effective model parameters for solute transport (e.g., *Porta et al.*, 2016; *Hochstetler and Kitanidis*, 2013; *Sanchez-Vila et al.*, 2010) since concentration profiles are a typical experimental observable (see for example *Gramling et al.*, 2002; *Berkoviz et al.*, 2000; *Ye at al.*, 2015; *Molins et al.*, 2014). We employ *OBF* in the maximum likelihood framework (*Carrera and Neuman*, 1986) to estimate $L$ and $R_D$ considering the concentration profile at $\hat{t}$ = 100 s for only the S_HV initial condition. This choice is consistent with the common practice of characterizing transport parameters in scenarios where the solute is injected in the porous domain in a flux-weighted fashion. The estimated values of $L$ and $R_D$ and corresponding variability inferred from the estimated standard deviations (as given by the approximated parameter covariance matrix, see *Carrera and Neuman*, 1986) are reported in Table 2.

Table 2: Best estimated values of $R_D$ and $L$ with the corresponding interval of confidence obtained by implementing *OBF* as optimization criterion for the calibration procedure applied to pore-scale average concentration data at $\hat{t}$ = 100 s and S_HV.

| Parameter | Estimated value | Interval of confidence |
|---|---|---|
| $L$ [μm] | 743 | [673 , 819] |
| $R_D$ | $10^{-1.011}$ | $[10^{-1.3257}, 10^{-0.6963}]$ |

As an alternative to *OBF*, we define *SOOF* as

$$P1 = \left| 1 - \frac{\gamma_F}{\gamma_{PS}} \right| \tag{14}$$

$$P2 = \left| 1 - \frac{\sigma^2{}_M}{\sigma^2{}_{PS}} \right| \tag{15}$$

$$SOOF = P1 + P2 \tag{16}$$

$\gamma_F$ and $\gamma_{PS}$ represent the skewness computed from the double-continuum model and pore-scale model implementation, respectively, at $\hat{t}$ = 400s considering the S_U initial condition. Quantities $\sigma^2{}_M$ and $\sigma^2{}_{PS}$ indicate the spread yielded by the continuum model and pore-scale model, respectively, at $\hat{t}$ = 50s considering the S_HV initial scenario. *SOOF* is specifically chosen based on the result of our sensitivity analysis results (*Razavi and Gupta,* 2016; *Pianosi et al.*, 2016). Based on the Sobol' sensitivity indices in Figure 6, we highlight that $\sigma^2$ in the S_HV scenario shows a substantial sensitivity to $L$, especially for early time ($\hat{t}$ = 10 s and 50 s) while $\gamma$ presents a marked sensitivity to $R_D$ at large time. This is corroborated by further inspection of the evolution of $P1$ and $P2$ through the parameter spaces $\Omega_L$ (Figure 7a) and $\Omega_{RD}$ (Figure 7b), respectively.

In Figure 7a and 7b we depict the evolution of $P1$ and $P2$ respectively for all combinations of parameters explored in the sensitivity analysis. The trend of $P1$ clearly suggests the presence of a minimum on the interval *IL*, i.e., *IL* = [ 763 μm, 863 μm]. Note that the confidence interval and the best estimate reported in Table 2 are in agreement with this interval *IL*. $P1$ shows steep gradients outside *IL* which is a desired feature of the objective function when dealing with the calibration process. In Figure 7b the $P2$ is displayed as function of $R_D$. $P2$ approaches its minimum in the interval *IRD* = $[10^{-2.37}, 10^{-1.955}]$. Note that the best estimate and the confidence interval of $R_D$ in Table 2 do not correspond to the indications given by $P2$ criterion. This is further discussed in Section 5.2. Similar to criterion $P1$, $P2$ shows a unique minimum located in a delimited area of the parameter space.

Criterion *SOOF* combining the quantities $P1$ and $P2$, results in an objective function sensitive to both parameters included in the continuum model. By observing the value of *SOOF* over the parameter space investigated in the sensitivity analysis (Figure 7c), we note that it has a global minimum corresponding to the intersection of the two white solid lines. *SOOF* is minimized using the *fmincon* Matlab function leading to $L$=673.4 μm and $R_D = 10^{-1.9972}$ which lie in *IL* and *IRD*, respectively.

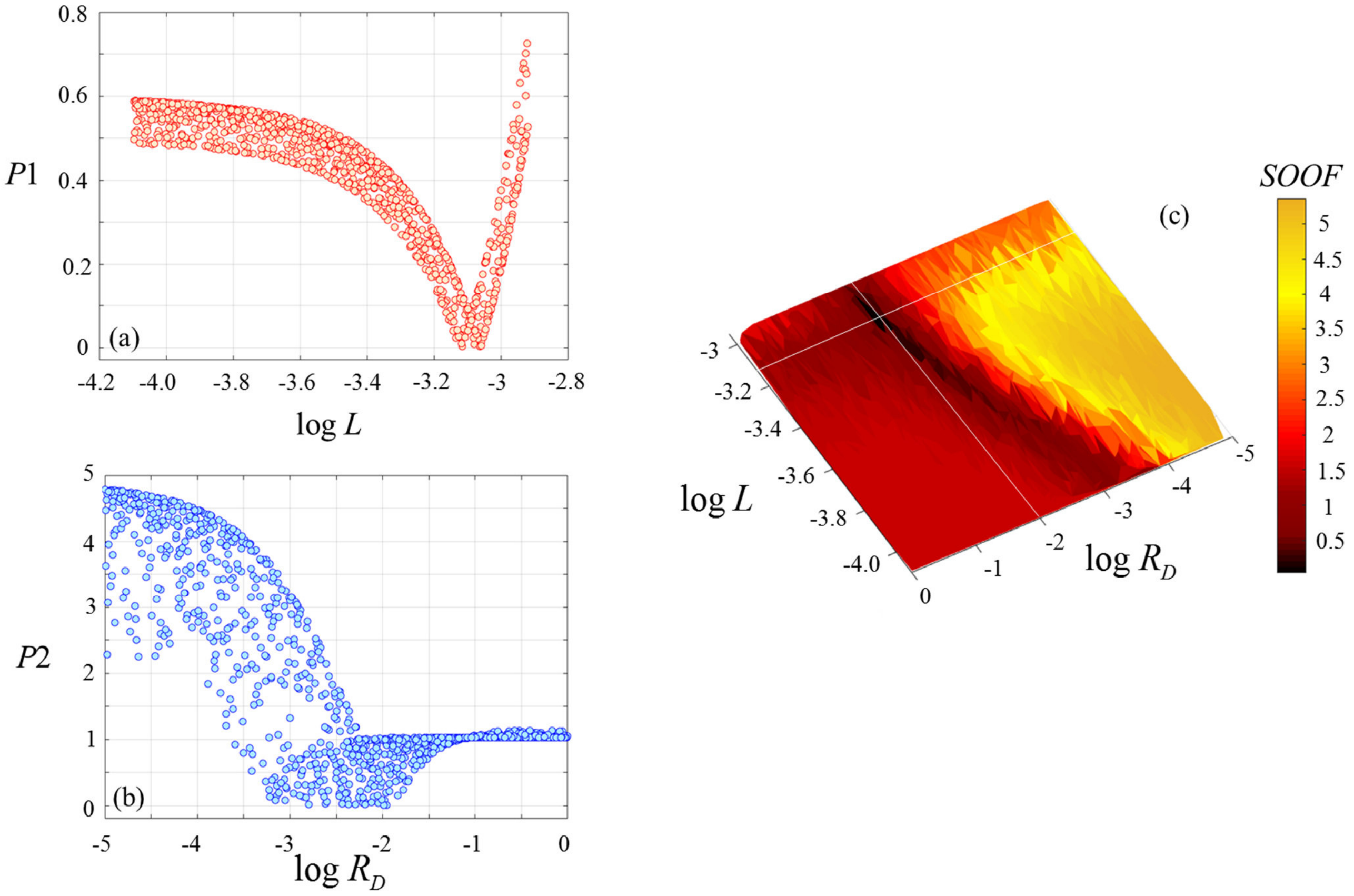

Figure 7: (a) Evolution of $P1$ as function of log $L$ and (b) evolution of $P2$ as function of log $R_D$ computed for the $N$ combinations of parameter explored in the sensitivity analysis. The shadowed (red and blue) areas represent the interval of parameters where the $P1$ and $P2$ quantities approach the minimum value. Panel (c) presents the evolution of *SOOF* throughout the parameter space explored in the sensitivity analysis. The solid white lines indicate the likely location of the minimum of the *SOOF*.

## 5.2 Model validation and discussion

In Figure 8, we present the comparison between the solute concentration profiles yielded by the model and the section averaged pore-scale data $E^*$ at different times ($\hat{t}$ = 50, 100, 200 and 400 s). The dashed lines represent the continuum model results using $L$ = 743 μm and $R_D = 10^{-1.011}$ based on *OBF* as discussed in Section 5.1. The solid lines indicate the continuum model profiles obtained for $L$=673.4 μm and $R_D = 10^{-1.9972}$ based on *SOOF*. Figure 8 a-c corresponds to the three different initial conditions, i.e. S_HV (Figure 8a), S_U (Figure 8b) and S_LV (Figure 8c). For simplicity we refer to M1 and M2 as predictions by the continuum model calibrated according to *OBF* and *SOOF*, respectively.

In Figure 8a, both M1 and M2 demonstrate good performance for scenario S_HV. The predictions of M1 and M2 are very close, leading to similar values of concentration peak, spread and shape for all the times investigated. A closer inspection reveals that at $\hat{t}$ = 100 s M1 better matches the pore-scale profile, consistent with the definition of *OBF*. At larger times ($\hat{t}$ = 200s and $\hat{t}$ = 400 s) a qualitative analysis suggests that M1 tends to match closely the frontward tail while M2 shows more accuracy in reproducing the backward tail. We avoid quantitative comparison given the intense fluctuations from the pore-scale data, which may yield misleading results.

The formulation of the continuum model presented in Sections 3 and 4 aims to embed the effect of pore-scale processes in the two effective parameters $L$ and $R_D$, which should depend only on pore structure and geometry. If so, in principle, the estimated parameters relying on *OBF* should be exportable to prediction of solute concentrations obtained in S_LV and S_U, even though these scenarios were not part of the calibration process. For S_U (Figure 8b), M1 provides a good interpretation of the forward tail and the concentration peak, but tends to underestimate the backward tail. This is clear especially at $\hat{t}$ = 100s and $\hat{t}$ = 200s. Concerning S_LV, M1 poorly interprets pore-scale data: the backward tail is markedly underestimated while the concentration peak is overestimated for all considered times shown in Figure 8c. Model M1 only presents a good match with data associated with fast flowing solute, i.e. a small portion of the frontward concentration tail. This is because M1 is calibrated to describe the mobile solute as shown in Figure 8a.

Based on the results in Figure 8, we can assert that the classical *OBF* allows calibration of a model with enough accuracy to predict concentration profiles at all times when the same initial condition is explored.

However we show that *OBF* may not be appropriate to estimate parameter values intended to be implemented for diverse initial conditions.

The combination of effective parameters (i.e. $L$=673.4 μm and $R_D = 10^{-1.9972}$) associated with M2 leads to improved predictions of the profiles for S_U and S_HV when compared to M1. In Figure 8b, M2 and M1 are virtually indistinguishable at $\hat{t}$ = 50s, but M2 better captures the behavior of the backward tail at longer times. This result is obtained by explicitly incorporating information on the skewness of the profile, which carries the necessary information to constrain $R_D$ (Figure 6). As the length scale in M1 ($L$ = 743 μm) is close to that of M2 ($L$=673.4 μm), we expect that the improved interpretation of the backward tails in Figure 8b is mainly attributable to the different order of magnitude in the estimate of $R_D$. Concerning the S_LV scenario, M2 shows excellent performance in reproducing the larger times ($\hat{t}$ = 200, 400 s) and matching the peak of solute concentration at $\hat{t}$=100 s even as it fails in capturing the exact profile shape at early times, i.e. $\hat{t}$ =100s and $\hat{t}$ =50s (Figure 8c). The mismatch at early times can be attributed to a limitation of the continuum model, which condenses the escape process from cavities with a single characteristic time scale $t_e$. This simplifies implementation, but only captures the average behavior of the cavities and not the complete distribution of transit times that may be present (see e.g., *Wirner et al.*, 2014). This limitation is highlighted when considering scenario S_LV where the escape process is crucial for determining concentration evolution.

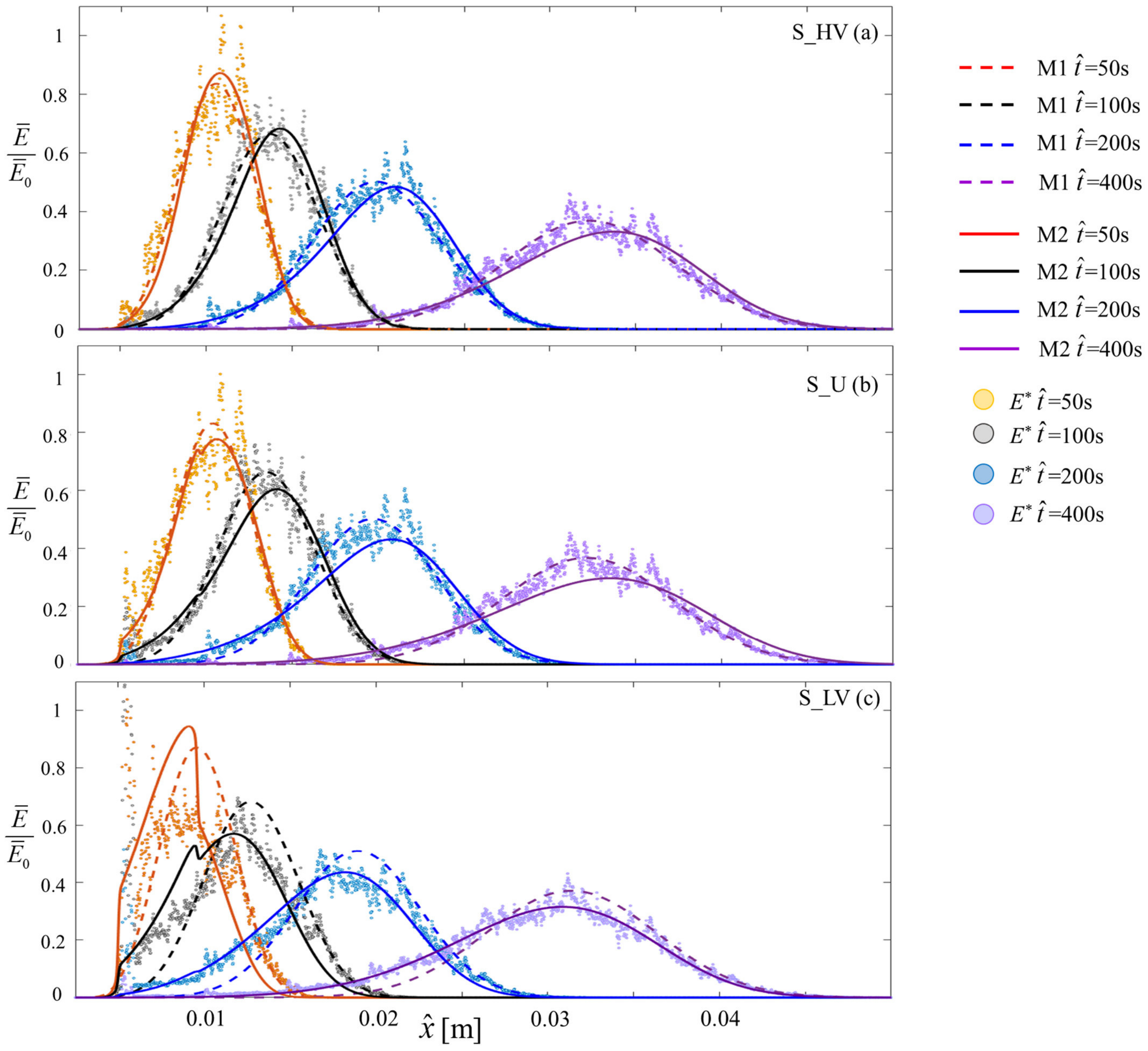

Figure 8: Comparison of concentration profiles yielded by M1 (dashed lines), M2 (soild lines) and pore-scale model (cricles) at $\hat{t}$ = 50, 100, 200 and 400s associaed with (a) S_HV, (b) S_U and (c) S_LV scenarios.

As discussed in Section 5.1, *SOOF* is tailor-made for this specific model and problem setting based on the sensitivity analysis in Section 4. By investigating the sensitivity, we can explain discrepancies and the reliability of parameter estimates obtained by using *OBF* compared to *SOOF*.

We conclude our analysis by a close inspection of the shortcomings in the implementation of *OBF*. Figure 9 shows that *OBF* has its minimum in the region, highlighted by the white rectangle. Criterion *OBF* sharply varies with $L$ close to the minimum, but has negligible gradients along $R_D$. This observation suggests that *OBF* is a strong criterion for estimating $L$. Indeed, the estimated values of $L$ from *OBF* and *SOOF* are consistent and reinforce one another. On the other hand, the identification of an optimal value for $R_D$ is problematic through *OBF*. Thereby, backward tails are not matched by M1. This result is in apparent contrast with the observation that the skewness ($\gamma$) of the profile obtained for S_HV is largely influenced by $R_D$ at $\hat{t}$ = 100 s (as indicated by the Sobol' indices $SI(\gamma)_{RD}$=0.43, see Figure 6), which would suggest that these data are appropriate to properly estimate $R_D$. By looking at Figure 9, we can note that the sensitivity of *OBF* to $R_D$ is not uniform across the parameter space for values of $L$ larger than 500 μm. This result agrees with previous observations that global sensitivity measures are effective in identifying general trends but may overlook local sensitivities, which are eventually relevant in a parameter estimation context (see also, *Ceriotti et al.*, 2018).

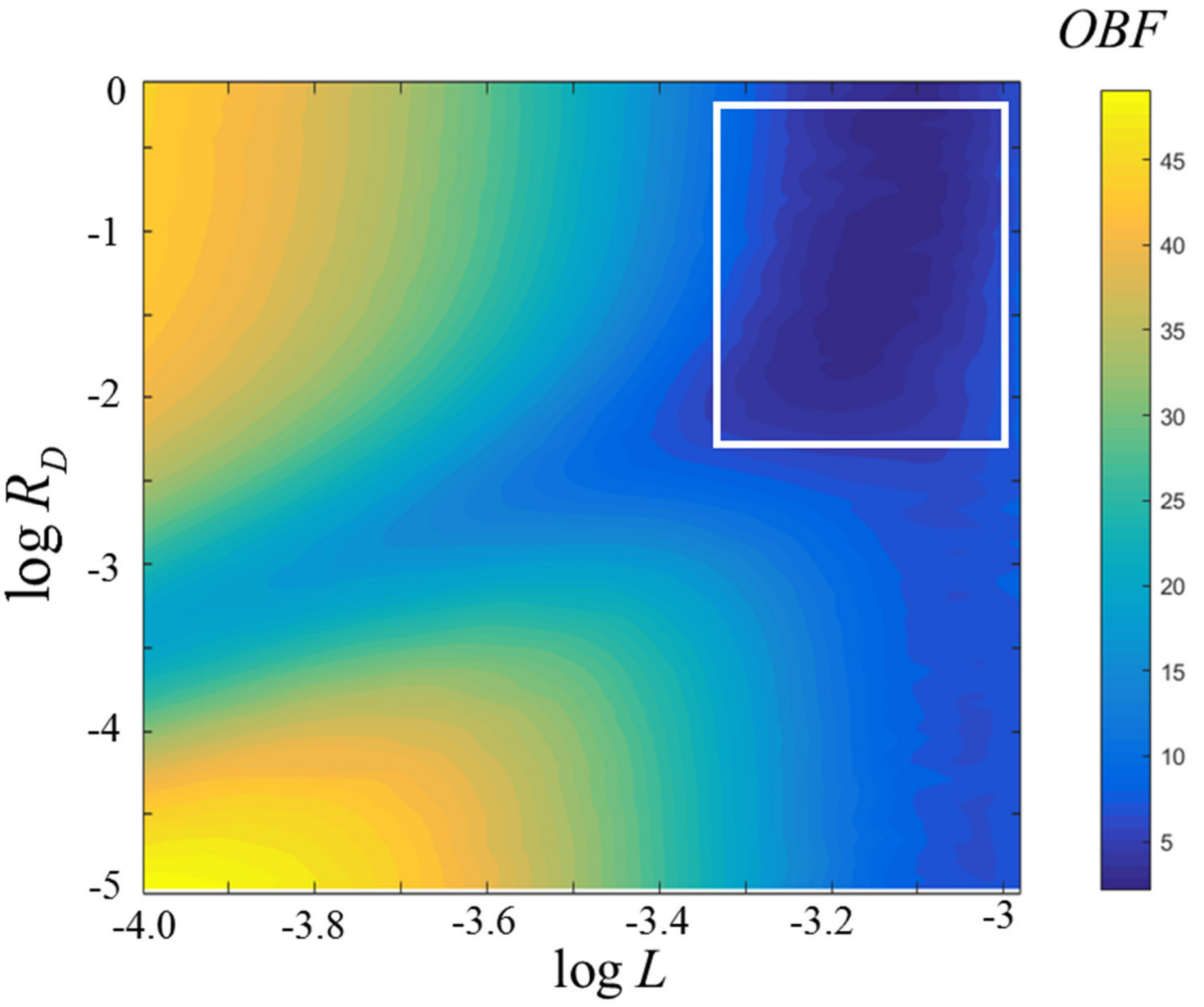

Figure 9: Evolution of the *OBF* across the parameter space. The white rectangle indicates the range of the parameter space where the *OBF* approaches the smallest values in the parameter space explored.

# 6 Conclusions

The present work is devoted to the formulation, calibration and validation of a double-continuum model for solute transport in porous media, which aims to embed characteristics of the pore-scale geometry and velocity field. Briefly, our methodology consists of: i) distinguishing high velocity and low velocity regions by comparing local pore-scale velocities to a fixed threshold value; ii) building a simplified unit cell made up of two separate immobile and mobile continua that can exchange mass; iii) characterizing the velocity profile in the mobile continuum through the cumulative density function of the local velocity field; iv) modeling the effect of complex structures such as cavities and dead end pores that can trap solute mass as a delayed diffusion process represented by a delay factor $R_D$ which controls the rate of mass exchange between the two continua. The resulting effective model embeds the effect stagnant regions in two parameters: $L$, a characteristic length scale and $R_D$.

After an efficient calibration, our proposed model is able to capture many crucial features of the concentration profile evolution over time for all considered scenarios (see Figure 8). Thus, despite the simplicity of model structure and the absence of spatial and temporal non local terms that arise in rigorous upscaling approaches, the model shows promising performances, yielding good fits and predictive capabilities to the diverse solute transport case studies considered here. We are able to identify a unique combination of effective parameters that characterize well, on average, the exchange process based solely on the features of the porous medium and the velocity field. However, its identification is not trivial and depends on the type of calibration data available and the structure of the objective function to be optimized. The use of classical mismatch between data and observations (*OBF*) for model calibration provides parameter estimates that perform well in describing solute transport for the S_HV scenario but these estimates were not exportable to other initial scenarios. In our case study, using a criterion driven by sensitivity analysis results (*SOOF*) for parameter estimation instead of *OBF* leads to an improved characterization of the model effective parameters, especially $R_D$. This result aligns with the idea that formulating an efficient objective function is a crucial aim along with the model development/implementation depending on the data available and the reference scenario investigated. This clearly suggests that the sensitivity analysis of an effective model can be an efficient tool to design *ad hoc* metrics for model calibration and circumvent the use of weakly sensitive objective functions for parameter estimation.

Moreover, investigating the role played by the different parameters by means of sensitivity analysis allows better understanding of the solute transport phenomena itself. Our results demonstrate that the impact of the exchange process on the solute profile is significant and can lead to notably different shapes and spreading patterns even when starting from the same initial condition (Figure 4). Modeling the access/escape process of stagnant zones is crucial for capturing the complex patterns associated with complex real porous media structures, e.g. such as those observed in carbonates.

Our analysis on three different initial conditions clearly suggests that the mass exchange process does not always affect the solute profile evolution: it is possible to observe a perfectly symmetric profile even in porous media characterized by extremely inaccessible cavities and dead end pores depending on the initial condition selected (see Figure 4b, S_HV, dashed line). This is clearly shown by the Sobol' indices results (Table 1 and Figure 6) which indicate that the exchange mass process and model parameters have a different impact depending on the initial scenario considered. We can conclude that tracer tests performed for a single initial condition may lead to incomplete knowledge of the porous medium structure and understanding of the impact of mass exchange processes.

Finally, we highlight that the proposed model is highly flexible and able to yield to symmetric profiles, highly skewed profiles or entrapped solute that can be retained for extremely long times (see Figure 3b and Figure 4). This motivates further developments by extending model testing to different real 3D porous media and exploring the existing specific relationship between porous geometry structure and effective parameters $L$ and $R_D$ which will be pursued in future works.

Acknowledgements

G. Ceriotti and G.M. Porta would like to thank the EU and MIUR for funding, in the frame of the collaborative international Consortium (WE-NEED) financed under the ERA-NET WaterWorks2014 Co- funded Call. This ERA-NET is an integral part of the 2015 Joint Activities developed by the Water Challenges for a Changing World Joint Programme Initiative (Water JPI).

D. Bolster greatly acknowledges financial support from the U.S. National Science Foundation via grants EAR 1351625 and CBET 1705770

Supporting Information for

# A double-continuum transport model for segregated porous media: derivation and sensitivity analysis-driven calibration

by G. Ceriotti[1], A. Russian[1], D. Bolster[2], G.M. Porta[1]

[1]Dipartimento di Ingegneria Civile e Ambientale, Politecnico di Milano, Piazza L. Da Vinci 32, 20133 Milano, Italy

[2]Department of Civil and Environmental Engineering and Earth Sciences, University of Notre Dame, Notre Dame, IN, USA.

## Analytical derivation of upscaled equations (4)-(6)

We describe here the procedure to upscale longitudinal transport and in particular to derive expressions (4)-(6). The procedure builds upon that of Porta et al. (2015). We start from the dimensionless system of equations (1)-(2), which are expressed in terms of concentrations in the mobile and immobile regions of the elementary cell illustrated in Figure 2a. Our aim is to obtain an upscaled effective longitudinal model written in term of the section-averaged concentrations:

$$\bar{E}_M = \frac{L}{l}\int_{-l/2L}^{l/2L} E_M dy; \quad \bar{E}_I = \frac{L}{L-l}\left(\int_{-1/2}^{-l/2L} E_I dy + \int_{l/2L}^{1/2} E_I dy\right) \tag{S1}$$

To this end we mimic the standard procedure employed to obtain Taylor-Aris dispersion between two parallel plates (Wooding, 1960).We start by averaging the two equations (1) along the $y$-direction of the unit cell.

$$\begin{cases} \dfrac{\partial \bar{E}_M}{\partial t} + \bar{u}\dfrac{\partial \bar{E}_M}{\partial x} = \dfrac{1}{Pe}\dfrac{1}{\tau_M}\dfrac{\partial^2 \bar{E}_M}{\partial x^2} - \dfrac{\partial \overline{\tilde{u}\tilde{E}_M}}{\partial x} + \dfrac{2}{Pe}\dfrac{1}{l}\left.\dfrac{\partial \tilde{E}_M}{\partial y}\right|_{y=l/2L} \\ \dfrac{\partial \bar{E}_I}{\partial t} = \dfrac{1}{Pe}\dfrac{1}{\tau_{IM}}\dfrac{\partial^2 \bar{E}_I}{\partial x^2} - \dfrac{2R_D}{Pe}\dfrac{1}{L-l}\left.\dfrac{\partial \tilde{E}_I}{\partial y}\right|_{y=l/2L} \end{cases} \tag{S2}$$

where

$$\tilde{u}(y) = u(y) - \frac{L}{l}\int_{-l/2L}^{+l/2L} u(y)\, dy \tag{S3}$$

is the deviation of the assumed velocity distribution (the cdf of the velocity modulus in the cell) with respect to its average value. Subtracting the section-averaged equations from (1) we obtain equations for the fluctuations

$$\begin{cases} \dfrac{\partial \tilde{E}_M}{\partial t} + \bar{u}\dfrac{\partial \tilde{E}_M}{\partial x} + \tilde{u}\dfrac{\partial \bar{E}_M}{\partial x} = \dfrac{1}{Pe\,\tau_M}\left(\dfrac{\partial^2 \tilde{E}_M}{\partial x^2} + \dfrac{\partial^2 \tilde{E}_M}{\partial y^2}\right) + \dfrac{\partial \overline{\tilde{u}\tilde{E}_M}}{\partial x} - \dfrac{2}{Pe}\dfrac{\phi}{\phi_{HV}}\left.\dfrac{\partial \tilde{E}_M}{\partial y}\right|_{y=l/2L} & 0 \le |y| < l/2L \\ \dfrac{\partial \tilde{E}_I}{\partial t} = \dfrac{1}{Pe\,\tau_{IM}}\left(\dfrac{\partial^2 \tilde{E}_I}{\partial x^2} + \dfrac{\partial^2 \tilde{E}_I}{\partial y^2}\right) + \dfrac{2R_D}{Pe}\dfrac{\phi}{\phi_{LV}}\left.\dfrac{\partial \tilde{E}_I}{\partial y}\right|_{y=l/2L} & l/2L < |y| < 1/2 \end{cases} \tag{S4}$$

with boundary conditions

$$\begin{aligned} &\frac{\partial \tilde{E}_M}{\partial y} = R_D \frac{\partial \tilde{E}_I}{\partial y} \quad ; \quad \tilde{E}_M = \tilde{E}_I - \Delta \bar{E}_{MI} && |y| = l/2L \\ &\frac{\partial \tilde{E}_I}{\partial y} = 0 && |y| = 1/2 \end{aligned} \tag{S5}$$

where $\Delta \bar{E}_{MI} = \bar{E}_M - \bar{E}_I$.

Following classical Taylor-Aris theory, we simplify equation (S4) and neglect gradients along $x$ relative to those along $y$, as well as time derivative terms, assuming a quasi-steady asymptotic state. We then obtain the following system of ordinary differential equations in $y$:

$$\begin{cases} \tilde{u}\dfrac{\partial \bar{E}_M}{\partial x} = \dfrac{1}{Pe}\dfrac{1}{\tau_M}\left(\dfrac{d^2 \tilde{E}_M}{dy^2}\right) - \dfrac{2}{Pe}\dfrac{\phi}{\phi_{HV}}\left.\dfrac{d\tilde{E}_M}{dy}\right|_{y=l/2L} & 0 \le |y| < l/2L \\ 0 = \dfrac{1}{Pe\,\tau_{IM}}\left(\dfrac{d^2 \tilde{E}_I}{dy^2}\right) + \dfrac{2R_D}{Pe}\dfrac{\phi}{\phi_{LV}}\left.\dfrac{d\tilde{E}_I}{dy}\right|_{y=l/2L} & l/2L > |y| > 1/2 \end{cases} \tag{S6}$$

Starting from (S5)-(S6), and relying on linear superposition, we can write the following closure expression

$$\tilde{E}_M(x,y) = b_1(y)\frac{\partial \bar{E}_M}{\partial x} + b_3(y)\Delta \bar{E}_{MI} \tag{S7}$$

$$\tilde{E}_I(x,y) = b_2(y)\frac{\partial \bar{E}_M}{\partial x} + b_4(y)\Delta \bar{E}_{MI} \tag{S8}$$

where $b_i(y)$ ($i$ =1 … 4) are closure variables that are used to quantify fluctuations with respect to average concentrations. Substituting (S7)-(S8) into (S5)-(S6) we find the following systems of equations

$$
\begin{cases}
\tilde{u} = \frac{1}{Pe}\frac{1}{\tau_M}\frac{d^2 b_1}{dy^2} - \frac{2}{Pe}\frac{\phi}{\phi_{HV}}\left.\frac{db_1}{dy}\right|_{y=l/2L} & 0 \le |y| < l/2L \\
0 = \frac{1}{Pe\tau_{IM}}\frac{d^2 b_2}{dy^2} + \frac{2R_D}{Pe}\frac{\phi}{\phi_{LV}}\left.\frac{db_2}{dy}\right|_{y=l/2L} & l/2L > |y| > 1/2 \\
b_1 = b_2;\ \frac{db_1}{dy} = R_D\frac{db_2}{dy} & |y| = l/2L \\
\frac{db_2}{dy} = 0 & |y| = 1/2
\end{cases}
\tag{S9}
$$

$$
\begin{cases}
0 = \frac{1}{Pe}\frac{d^2 b_3}{dy^2} - \frac{2}{Pe}\frac{\phi}{\phi_{HV}}\left.\frac{db_3}{dy}\right|_{y=l/2L} & 0 \le |y| < l/2L \\
0 = \frac{R_D}{Pe}\frac{d^2 b_4}{dy^2} + \frac{R_D}{Pe}\frac{\phi}{\phi_{LV}}\left.\frac{db_4}{dy}\right|_{y=l/2L} & l/2L < |y| < 1/2 \\
b_3 = b_4 - 1; \frac{db_3}{dy} = R_D\frac{db_4}{dy} & |y| = l/2L \\
\frac{db_4}{dy} = 0 & |y| = 1/2
\end{cases}
\tag{S10}
$$

The two systems (S9)-(S10) are not coupled and can be solved independently. System (S9) needs to be solved numerically, if $\tilde{u}(y)$ does not have a close form analytical expression. To this end, we employ a standard finite difference approach. System (S10) can be solved analytically once values for $\phi_{HV}$, $\phi_{LV}$ are determined and $R_D$ is fixed.

$b_i(y)$ they can then be replaced in the section-averaged system (S2), by (S7)-(S8), to obtain the longitudinal transport model in (4) which is reported here for completeness.

$$
\begin{cases}
\frac{\partial \bar{E}_M}{\partial t} + \frac{\partial \bar{E}_M}{\partial x} + \frac{\partial}{\partial x}\left[d_{H1}\frac{\partial \bar{E}_M}{\partial x} + d_{H2}\Delta\bar{E}_{MI}\right] = \frac{1}{Pe}\frac{1}{\tau_M}\frac{\partial^2 \bar{E}_M}{\partial x^2} + \frac{\phi}{Pe\phi_{HV}}\left(e_1\frac{\partial \bar{E}_M}{\partial x} + e_2\Delta\bar{E}_{MI}\right) \\
\frac{\partial \bar{E}_I}{\partial t} = \frac{1}{\tau_{IM}}\frac{1}{Pe}\frac{\partial^2 \bar{E}_I}{\partial x^2} - \frac{\phi}{\phi_{LV}Pe}\left(e_1\frac{\partial \bar{E}_M}{\partial x} + e_2\Delta\bar{E}_{MI}\right)
\end{cases}
\tag{S11}
$$

with

$$
d_{H1} = \frac{L}{l}\int_{-l/2L}^{l/2L} b_1\tilde{u}\ \mathrm{d}y;\quad d_{H2} = \frac{L}{l}\int_{-l/2L}^{l/2L} b_3\tilde{u}\ \mathrm{d}y \tag{S12}
$$

$$
e_1 = 2R_D\left.\frac{\partial b_2}{\partial y}\right|_{y=l/2L} = 2\left.\frac{\partial b_1}{\partial y}\right|_{y=l/2L};\quad e_2 = 2R_D\left.\frac{\partial b_4}{\partial y}\right|_{y=l/2L} = 2\left.\frac{\partial b_3}{\partial y}\right|_{y=l/2L} \tag{S13}
$$

In the application presented in this work the values of $\phi_{HV}$, $\phi_{LV}$ are fixed a priori, as explained in section 2. The procedure employed to obtain each solution of the upscaled model is the following:

1. Fix the values of parameters $L$ and $R_D$
2. From $L$ compute $Pe$
3. Compute the closure variables $b_i(y)$ from (S9)-(S10)
4. Quantify upscaled parameters through (S12)-(S13)
5. After fixing the appropriate initial condition (see Section 2.2), simulate longitudinal transport by numerically solving system (S11) (here performed through the Matlab function pdepe).

In this study we implemented the procedure in Matlab (MATLAB ® and Statistics Toolbox Release 2016b). Overall performing the five steps above requires a computational time of ~60s on an Intel(R) Core(TM) i7-47110HQ CPU @ 2.50 GHz and 16 Gb RAM.